\documentclass{moriond}

\usepackage{xspace}
\usepackage{amsmath}
\def\Journal#1#2#3#4{{#1} {\bf #2}, #3 (#4)}

\def\be{\begin{equation}}
\def\ee{\end{equation}}
\def\bea{\begin{eqnarray}}
\def\eea{\end{eqnarray}}

\newcommand{\Photo}{}

\newcommand{\chisq}{\ensuremath{\chi^{2}}\xspace}

\newcommand{\mjj}{\ensuremath{m_\text{jj}}\xspace}

\newcommand{\ys}{\ensuremath{y^*}\xspace}
\newcommand{\yb}{\ensuremath{y_\text{b}}\xspace}

\newcommand{\mz}{\ensuremath{m_\text{Z}}\xspace}
\newcommand{\as}{\ensuremath{\alpha_\text{s}}\xspace}
\newcommand{\asmz}{\ensuremath{\as(\mz)}\xspace}
\newcommand{\ptavg}{\ensuremath{\langle{}p_\text{T}\rangle_{1,2}}\xspace}
\newcommand{\ndf}{\ensuremath{n_\text{dof}}\xspace}

\begin{document}
\vspace*{4cm}
\title{Precision determination of $\alpha_s$ from Dijet Cross Sections in the Multi-TeV Range}

\author{ Jo\~ao Pires$^{1, 2}$ \footnote[2]{Speaker}}

\address{
$^{1}$LIP, Avenida Professor Gama Pinto 2, P-1649-003 Lisboa, Portugal\\
$^{2}$Faculdade de Ci\^encias, Universidade de Lisboa, 1749-016 Lisboa, Portugal
}

\maketitle\abstracts{
In this talk we present a determination of the strong coupling constant $\alpha_s$ and its energy-scale dependence based on a next-to-next-to-leading order (NNLO) QCD analysis of dijet production. 
Using the invariant mass of the dijet system to probe $\alpha_s$ at different scales, we extract a value of $\alpha_s(m_Z) = 0.1178\pm0.0022$ from LHC dijet data. The combination of various LHC datasets significantly extends the precision 
and scale reach of the analysis, enabling the first determination of $\alpha_s$ up to 7 TeV. By incorporating dijet cross sections from HERA, we further probe $\alpha_s$ at smaller scales, covering a kinematic range of 
more than three orders of magnitude. Our results are in excellent agreement with QCD predictions based on the renormalization group equation, providing a stringent test of the running of the strong coupling across a wide energy range.
}

\section{Introduction}
In Quantum Chromodynamics (QCD), apart from the quark masses, the strong coupling constant $\alpha_s$ is the only free parameter of the QCD Lagrangian. It governs the strength of the interaction between 
quarks and gluons and enters into the calculation of every process involving the strong interaction. Despite its fundamental role, $\alpha_s$ remains the least precisely known of the fundamental couplings, with a relative uncertainty of order 1\% 
($\delta \alpha_s / \alpha_s \sim 1\%$), compared to much smaller uncertainties for other couplings, such as ($\delta G_F / G_F \sim 10^{-5}$) and ($\delta \alpha / \alpha \sim 10^{-9}$). This limited precision in $\alpha_s$ propagates 
into theoretical uncertainties in precision analyses at the LHC, for example in Higgs boson cross sections and searches for new physics. 

While $\alpha_s$ at a fixed scale $Q$ must be determined experimentally, its scale dependence is predicted by QCD through the renormalization group equation, leading to asymptotic freedom: at high energies 
$\alpha_s$ becomes small and perturbative methods are applicable, while at low energies the coupling grows, signaling the onset of non-perturbative dynamics and confinement.

Dijet production at hadron colliders provides a direct and sensitive probe of $\alpha_s$ across a wide range of scales. In this talk, we present new determinations of $\alpha_s$ based on a next-to-next-to-leading order 
(NNLO) QCD analysis of dijet production data from the LHC and HERA experiments. The use of NNLO theory significantly reduces theoretical uncertainties, and the combination of 
LHC and HERA data allows us to test the running of $\alpha_s$ from low scales to the multi-TeV regime, providing a stringent test of QCD over more than three orders of magnitude in energy.
A full account of the analysis method presented in this talk can be found in~\cite{as}. Here, we summarize the essential points relevant for understanding the determination of $\alpha_s$ and its running.

\section{Methodology}
The objective function used in the fitting algorithm is derived from normally distributed relative uncertainties and defined as~\cite{chi}
\begin{equation}
  \chisq = \sum_{i,j}
  \log\frac{\varsigma_i}{\sigma_i}
  \left(V_\text{exp} + V_\text{NP} + V_\text{NNLOstat} + V_\text{PDF}\right)_{ij}^{-1}
  \log\frac{\varsigma_j}{\sigma_j}\,,
  \label{eq:chisq}
\end{equation}
where the double-sum runs over all data points, $\varsigma_{i}$ denotes the measured cross section, $\sigma_i$ denotes the theory prediction, and the minimization of \chisq with respect to $\alpha_s(m_Z)$ is performed
using TMinuit's Migrad algorithm~\cite{Migrad}. The experimental datasets $\varsigma_{i}$ comprises 
inclusive dijet cross section measurements from the ATLAS and CMS experiments at proton–proton center-of-mass energies of 7, 8, and 13 TeV. ATLAS provides double-differential cross sections as functions of the dijet invariant 
mass \mjj and half the rapidity separation $y^*=|y_1-y_2|/2$ at 7 and 13 TeV~\cite{ATLAS}. CMS reports double-differential measurements in $m_{jj}$ and the maximum absolute rapidity $y_\text{max}$ of the two leading-$p_T$ jets
at 7 and 13 TeV, as well as triple-differential cross sections at 8 and 13 TeV in terms of \mjj (or the average transverse momentum \ptavg), $y^*$, and the longitudinal boost $\yb = \left|y_1+y_2\right|/2$~\cite{CMS}. 
All measurements use the anti-$k_T$ jet algorithm, with the larger available jet size parameter $R$ selected to ensure better perturbative convergence.

The covariance matrices $V_\text{exp}$, $V_\text{NP}$, $V_\text{NNLOstat}$, and $V_\text{PDF}$ in~\eqref{eq:chisq} represent the relative experimental, non-perturbative (NP), NNLO statistical, and PDF uncertainties, respectively,
 as described in detail in~\cite{as}. Experimental uncertainties are typically at the few-percent level across most of the phase space. They are driven primarily by the jet energy scale (JES), with uncertainties ranging between 2\% 
 and 5\% at moderate \mjj, and reaching up to 30\% in the most extreme kinematic regions, such as forward rapidities, large rapidity separations, or highly boosted dijet systems. Statistical uncertainties remain small (generally below 5\% 
 in central regions).
 %, while contributions from jet energy resolution (JER), luminosity, trigger, and modeling effects are at the few-percent level, but can be larger in forward or boosted configurations and at large rapidity separation.
 The overall uncertainty estimates incorporate multiple systematic sources and statistical correlations arising from data unfolding. However, correlations between uncertainties across different datasets are not provided and thus treated as independent.
 
Non-perturbative correction uncertainties and PDF uncertainties are accounted for following the methods described in~\cite{as}. The NP uncertainties incorporate variations among event
generators and hadronization models with moderate bin-to-bin correlations assumed, while PDF uncertainties are obtained from the respective PDF set in the LHAPDF format~\cite{LHAPDF}. By considering them as a covariance 
matrix in \chisq, the PDF uncertainties are further constrained by the jet data. 

In turn, fixed-order predictions for the two- and three-dimensional dijet cross sections are computed at NNLO accuracy in perturbative QCD using the NNLOJET framework~\cite{NNLOJET}. 
The cross section is expressed through the QCD factorization theorem as
\[
\mathrm{d}\sigma = 
\sum_{a,b} 
\int \frac{\mathrm{d}x_1}{x_1} 
      \frac{\mathrm{d}x_2}{x_2} \,
      f_a(x_1, \mu_F, \alpha_s(\mu_F)) \,
      f_b(x_2, \mu_F, \alpha_s(\mu_F)) \,
      \mathrm{d}\hat{\sigma}_{ab}(\mu_R, \mu_F, \alpha_s(\mu_R)).
\]
where \( f_a(x, \mu_F, \alpha_s) \) are the parton distribution functions, and 
\( \mathrm{d}\hat{\sigma}_{ab} \) are the partonic cross sections calculated perturbatively. 
Both components depend on the strong coupling, \( \alpha_s \), through
the evolution of the PDFs and in the perturbative expansion of the matrix elements. In particular, the $x$-dependence of the PDFs $f_{x,\mu_0}$ is taken from PDF4LHC21~\cite{PDF4lhc} at 
a starting scale of $\mu_0=90$ GeV and is then evolved to the approriate renormalization ($\mu_R$) and factorization ($\mu_F$) scale $\mu=\mu_R=\mu_F=\mjj$,  the invariant mass of the dijet system.
This evolution uses the DGLAP equations, treating $\alpha_s(M_Z)$ as a free parameter. The evolved PDFs are given by
\begin{equation}
  f_a(x,\mu,\as) = ({\Gamma}(\mathcal{P},\mu,\mu_0,\as) \otimes f_{\mu_0})_a\,,
\end{equation}
where $\Gamma$ represents the DGLAP evolution kernels, evaluated at three-loop order~\cite{DGLAP} with the program Apfel++~\cite{APFEL}. As a consequence, two additional sources of theoretical uncertainty are considered in the 
determination of $\alpha_s$. The first is associated with the choice of the initial scale of the PDF evolution $(\mu_0)$, which is varied by factors of 0.5 and 2 around the nominal value.
The second source originates from the choice of renormalization $\mu_R$ and factorization $\mu_F$ scales, reflecting uncertainties from missing higher-order corrections and the arbitrariness of these scale choices. 
As this uncertainty reflects the sensitivity of the predictions to unphysical scale choices, it cannot be constrained by the data and is not included in 
the $\chisq$ minimization of the fit. It is estimated via independent variations of $\mu_R$ and $\mu_F$ by factors of 0.5, 1, and 2 around the nominal scale, using the standard 7-point variation.

In order to efficiently evaluate theoretical predictions for arbitrary PDF sets, $\mu_R$ and $\mu_F$ scales, 
and different values of \(\alpha_s(m_Z)\), as required in the fitting procedure, NNLOJET is interfaced with the APPLfast library~\cite{APPLFAST}. 
This approach uses interpolation grids, enabling fast recalculations without rerunning the full NNLO computation, and includes all subleading colour contributions present at NNLO. 
The predictions are further corrected for non-perturbative effects and higher-order electroweak contributions using correction factors as published by the experimental collaborations.

\begin{figure}
\centerline{\includegraphics[width=0.8\linewidth]{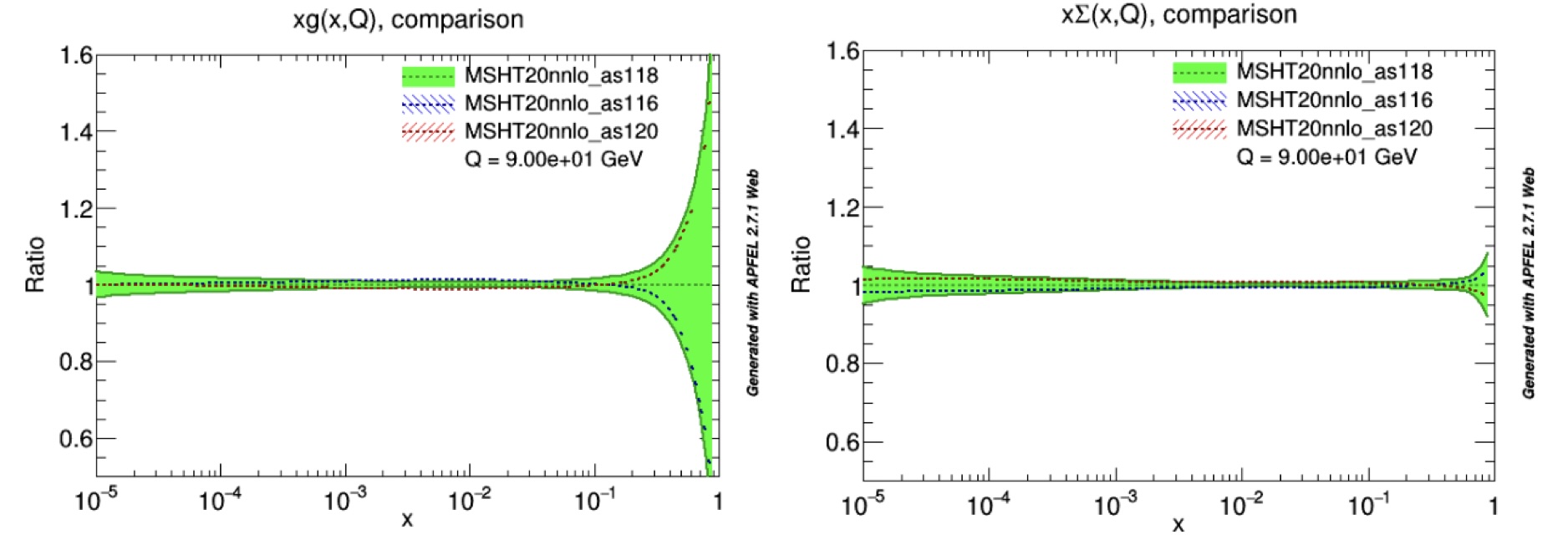}}
\caption[]{Plots from APFELweb showing gluon and Sigma PDFs from MSHT20nnlo at the nominal starting scale $\mu_0=90$ GeV, for different $\alpha_s(M_Z)$ assumptions. The green shaded areas represent PDF uncertainties.}
\label{fig:pdfsalpha}
\end{figure}

A potential concern in the methodology is that the initial PDF shape at the reference scale $\mu_0$ may itself depend on the value of $\alpha_s$ assumed in the original PDF fit. To assess this, we have explicitly checked the 
effect of varying $\alpha_s$ in a PDF set (cf. Fig~\ref{fig:pdfsalpha}). We find that the impact of this variation on the fitted $\alpha_s$ is comparable to the effect of varying the nominal starting scale of the PDF evolution 
by a factor of two around the central value and is effectively covered by it. This confirms that any residual bias from not refitting the PDFs at each $\alpha_s$ value is small and covered by our assigned theoretical uncertainty.
Moreover, predictions are compared to data using a full covariance matrix that includes both experimental and PDF uncertainties (accounting for $\alpha_s$ effects in the evolution), ensuring that the 
impact of PDF uncertainties on the extracted $\alpha_s$ is correctly propagated within the fit. To minimize sensitivity to PDF uncertainties and reduce moderate tensions observed in some phase space regions, 
we restricted the data used in the fit to measurements with $\ys<2.0$ and $\yb<1.0$~\cite{as}. This selection effectively restricts the PDFs $x$-range to $x>10^{-2}$. 
Altogether, 367 out of 493 cross section measurements are included in the nominal determination of $\alpha_s$ from LHC dijet data.

\begin{figure}
\centerline{\includegraphics[width=1.0\linewidth]{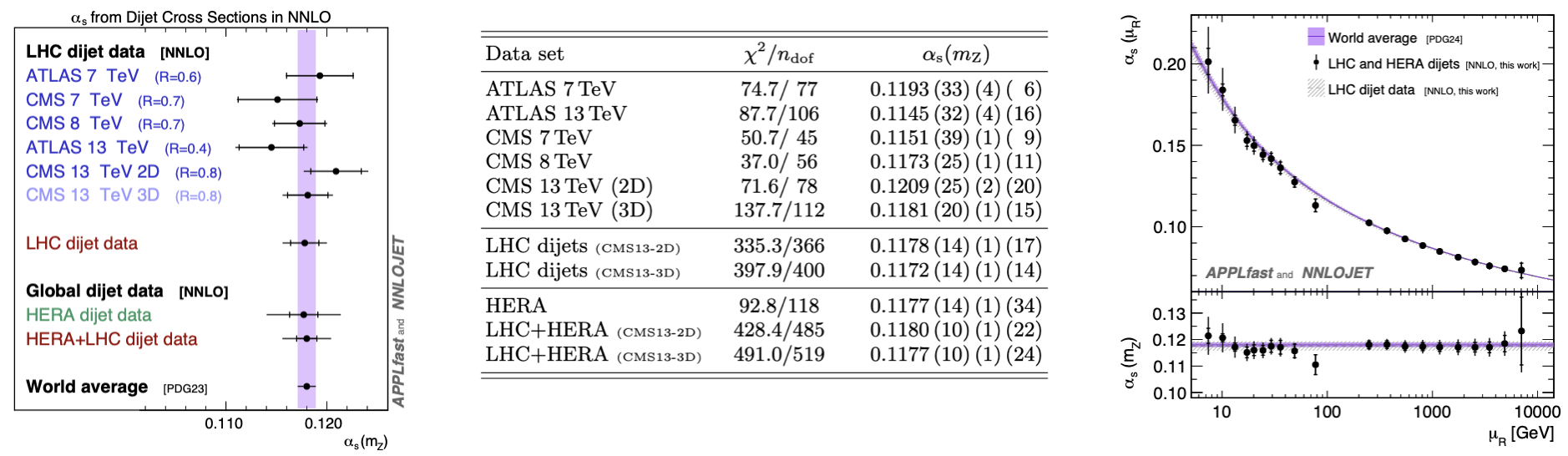}}
\caption[]{The left plot compares \asmz determinations from dijet cross sections to the world average. Inner error bars show (fit, PDF) uncertainties; outer error bars additionally include scale and $\mu_0$ uncertainties~\cite{as}. The center table 
summarizes \asmz values from fits of NNLO pQCD predictions to dijet cross section data, listing (fit, PDF), $\mu_0$, and scale $(\mu_R, \mu_F)$ uncertainties. The right plot shows the running of the strong coupling with the 
renormalization scale: the upper panel displays $\alpha_s(\mu_R)$, and the lower panel shows \asmz alongside the world average~\cite{asavg}. The hatched area indicates \asmz from LHC dijet data and its running with $\mu_R$~\cite{as}.}
\label{fig:asresults}
\end{figure}

\section{Results}
The strong coupling constant at the Z boson mass scale, $\alpha_s(m_Z)$, was determined using five LHC dijet datasets with full NNLO pQCD predictions, with the results reported in Fig.~\ref{fig:asresults}. 
The combined fit showed excellent consistency ($\chisq/\ndf=0.92)$ and yielded $\alpha_s(m_Z)=0.1178$ with uncertainties from fit and PDFs ($\pm0.0014$), scale $\mu_0$ ($\pm0.0001$), and scale variations $\mu_R,\mu_F$ ($\pm0.0017$). 
Individual datasets gave results consistent with the world average but with larger uncertainties ($\pm0.0020$ to $\pm0.0039$), which improved for datasets with higher luminosity and energy. 
The combined fit benefits from independent measurements, extended kinematic coverage, and multiple energies, reducing experimental uncertainties compared to single datasets.

Including dijet cross sections measured in electron-proton collisions at the HERA collider, we performed one of the most 
comprehensive and precise tests of the QCD renormalization group running of $\alpha_s(\mu)$. The running is probed through fits to individual \mjj ranges, and excellent agreement is found 
with the QCD prediction, as shown in Fig.~\ref{fig:asresults}. By combining HERA and LHC data, the behavior of the strong coupling as a function of energy is tested over an unprecedented range, 
from about 7 GeV to 7 TeV.

\section*{Acknowledgments}
The work presented in these proceedings has been carried out in collaboration with F. Ahmadova, D. Britzger, X. Chen, J. G{\"a}{\ss}ler, A. Gehrmann-De Ridder, T. Gehrmann, N. Glover, C. Gwenlan, 
G. Heinrich, A. Huss, L. Kunz, K. Rabbertz and M. Sutton. The work of JP is funded by Funda\c{c}\~{a}o para a Ci\^{e}ncia e Tecnologia (FCT-Portugal), through the programatic funding of R\&D units  
UIDP/50007/2020 and under projects CERN/FIS-PAR/0032/2021, 2024.05140.CERN.


\section*{References}
\begin{thebibliography}{99}

\bibitem{as} F. Ahmadova, D. Britzger, X. Chen et. al arXiv:2412.21165
\bibitem{chi} V. Andreev {\it et al}, (H1), \Journal{Eur. Phys. J. C}{75}{65}{2015}
\bibitem{Migrad} F. James and M. Roos, \Journal{Comput. Phys. Commun.}{10}{343}{1975} \\
			  R. Brun and F. Rademakers, \Journal{Nucl. Instrum. Meth.}{A389}{81}{1997}
\bibitem{ATLAS} (ATLAS), \Journal{JHEP}{2014}{05}{059}  (ATLAS), \Journal{JHEP}{2018}{05}{195}
\bibitem{CMS}    (CMS) \Journal{Phys. Rev. D}{87}{112002}{2013} (CMS) \Journal{Eur. Phys. J. C}{77}{11,746}{2017} \\
		          (CMS) \Journal{Eur. Phys. J. C}{865}{1,72}{2025} 
\bibitem{LHAPDF} A. Buckley {\it et al}, \Journal{Eur. Phys. J. C}{75}{132}{2015}
\bibitem{NNLOJET} A. Huss {\it et al}, arXiv:2503.22804
\bibitem{PDF4lhc}  R. D. Ball {\it et al}, (PDF4LHC Working Group) \Journal{J. Phys. G}{49}{080501}{2022}
\bibitem{DGLAP}  	A. Vogt, S. Moch, and J. A. M. Vermaseren, \Journal{Nucl. Phys. B}{691}{129}{2004} \\
				S. Moch, J. A. M. Vermaseren, and A. Vogt, \Journal{Nucl. Phys. B}{688}{101}{2004}
\bibitem{APFEL} 	V. Bertone, S. Carrazza, and J. Rojo \Journal{Comput. Phys. Commun.}{185}{1647}{2014}
\bibitem{APPLFAST} D. Britzger {\it et al}, \Journal{Eur. Phys. J. C}{82}{930}{2022} 
				 \Journal{Eur. Phys. J. C}{79}{845}{2019}
\bibitem{asavg} S. Navas et al. (Particle Data Group), \Journal{Phys. Rev. D}{110}{030001}{2024}

\end{thebibliography}
\end{document}